\documentclass[preprint,12pt]{elsarticle}

\usepackage{amssymb}

\newcommand{\revised}{}

\journal{Space Policy}

\begin{document}

\begin{frontmatter}

\title{A Model for Economic Freedom on Mars}

\author[inst1]{Jacob Haqq-Misra}
\ead{jacob@bmsis.org}

\affiliation[inst1]{organization={Blue Marble Space Institute of Science},Department and Organization
            addressline={600 1st Avenue, 1st Floor}, 
            city={Seattle},
            state={WA},
            postcode={98104}, 
            country={USA}}

\begin{abstract}
The momentum of human spaceflight initiatives continues to build toward Mars, and technological advances may eventually enable the potential for permanent space settlement. Aspirations for sustaining human life in space must be predicated on human factors, rather than technological constraints alone, and advances in models of governance and ethics are necessary as human civilization becomes a spacefaring species. This paper presents an idealistic but feasible model for economic freedom on Mars, which is situated within a framework in which Mars has been designated as a sovereign juridical peer to Earth. Under such conditions, Mars could maintain monetary stability through full reserve banking and a restriction on exchange with any fractional reserve Earth currencies, with a volume of circulating currency that changes based on the total population within fixed capacity infrastructure. Mars could maintain long-term political stability by diffusing the ownership of capital on Mars, which would allow all citizens of Mars to draw sufficient wealth from a combination of capital ownership and labor to live a good life. This model could also support limited tourism on Mars, in which real goods are exchanged for services but currency transactions between planets are prohibited. This model demonstrates the potential for a viable and sustainable economy on Mars that could conceivably be implemented, including on a sovereign Mars but also in other scenarios of space settlement. More broadly, this model illustrates that ideas such as diffuse capital ownership and limited government can enable freedom in space, and numerous models beyond a centralized world space agency should be explored to ensure the optimal governance of the emerging space economy.  
\end{abstract}



\end{frontmatter}

\newpage


\section{Introduction}
\label{sec:introduction}

Numerous national and private space are developing plans for sending humans to Mars, which includes ambitions for long-term settlement. The timelines for such missions are continually revised, but the continued investment in the technological capabilities for enabling long-duration human spaceflight will eventually enable the ability for travel to Mars. This momentum toward human exploration and settlement of the moon, Mars, and asteroids is driven by a multiplicity of stakeholders with various commercial, political, and scientific interests. The ultimate success of any attempt to inhabit space remains uncertain, but attempts will nevertheless be made in the coming decades to centuries. 

The challenges of technologically adapting to the space environment also raise parallel challenges for systems of ethics and governance. As ongoing research in the requisite technology continues, a community of scholars has begun to discuss the human dimensions of space settlement, drawing upon disciplines such as ethics, governance, history, planetary protection, sociology, psychology, and futures studies \citep[e.g.,][]{milligan2014nobody,szocik2019human,gilley2020space,froehlich2021assessing,cockell2022institutions,cockell2022interplanetary,haqq2022sovereign,nesvold2023off,schwartz2023reclaiming}. This ongoing effort is based on the recognition that space settlement is ultimately a human endeavor, and any aspirations for the long-term presence of human life in space must be predicated on human factors, rather than the assumption that technological capabilities are the only requirement. 

The idea of conceptualizing models of governence or ethical frameworks for application to martian settlements or the space environment is valuable as more than an academic exercise. The most apparent value is in the potential for applications to the long-term future of humans in space: any feasible models of governance or ethics that are developed today remain possibilities to draw upon when humans actually begin to live in space. The analysis by \citet{haqq2023constraints} argues that any model for space settlement, whether pragmatic or idealistic, should pass a three-prong test of its technological capability, political feasibility, and long-term sustainability in order be considered viable. Any such viable models that are developed in advance of human settlement can provide guidance, examples, or precedent for future developments in space. But a second source of value is gained by using space settlement as a venue for thought experiments in ethics and governance, which can yield novel insight about governance principles that may be taken for granted on Earth. The physical properties of the space environment impose firm limitations on human activities, and any theoretical governance models to address such challenges in space may also find areas of relevance on Earth. Even if Mars settlement is a long way off, or never occurs, the activity of thinking about governance in space remains useful for imagining the possibilities for governing our own future on Earth. 

This paper presents an idealistic, but feasible, model for economic freedom on Mars. This approach is specifically based on the idea of establishing Mars as a sovereign juridical peer to Earth, following the requirements developed by \citet{haqq2022sovereign} (discussed in \S\ref{sec:sovereignmars}). Within these requirements, this model suggests that long-term economic stability on Mars could be maintained by implementing a form of full-reserve banking, based on the ideas of \citet{douglas1939program} (discussed in \S\ref{sec:banking}). This model also suggests that long-term political stability could be aided by reducing inequalities in resource distribution by adopting a model of expanded capital ownership, based on the ideas of \citet{kelsoadler} (discussed in \S\ref{sec:capital}). Such a model could enable Mars to retain its sovereignty while also allowing a limited form of tourism on Mars (discussed in \S\ref{sec:tourism}). This model is intended to provide a feasible possibility for economic freedom on Mars that can stimulate further thinking about governance in space, and on Earth.

\section{Requirements for a Sovereign Mars}\label{sec:sovereignmars}

The conditions for establishing Mars as an independent planetary state, a juridicial peer to Earth were developed in the book \textit{Sovereign Mars} by \citet{haqq2022sovereign}. This book includes a historical analysis of cooperative sovereignty on Earth, using examples such as the Outer Space Treaty, the Convention on the Law of the Sea, and the Antarctic Treaty System to consider the future evolution of sovereignty in space. One result of this analysis is a working hypothesis for future governance models in space: (i) no world space agency (any new international organizations will not hold significant jurisdiction over spacefaring activities); (ii) no equitable sharing (mandatory equitable sharing of wealth will not gain complete participation from spacefaring states); and (iii) no new space treaties (any new treaties will be ineffective without participation by major spacefaring states). This working hypothesis could be falsified, for example through a major shift in geopolitics. But taken as a starting point, this working hypothesis can be used for imagining possibilities for governance in space that are at least conceptually situated within the realities of today.

Within the constraints of this working hypothesis, \citet{haqq2022sovereign} suggests an idealistic, although conceptually feasible, model for martian governance that would establish Mars as a sovereign planet prior to the arrival of the first humans. The purpose of such an endeavor would be to maximize the transformative potential of Mars to enable new experiments in civilization\revised{---in other words, the sovereign Mars model intentionally restricts the influence of Earth on Mars in order to allow for the greatest possible independence in the evolution of political and social systems on Mars. This is an admittedly idealistic approach that may be difficult to implement, but the sovereign Mars model is a framework that could enable---at least theoretically---the establishment of a second instance of civilization.} \citet{haqq2022sovereign} provides five specific requirements for establishing a sovereign Mars:
\begin{quote}
    \begin{enumerate}
        \item Humans who leave Earth to permanently settle on Mars relinquish their planetary citizenship as earthlings and claim a planetary citizenship as martians. This requires renunciation of any state or local citizenships. Martians cannot represent any interests of Earth and cannot acquire wealth on Earth.
        \item Earthlings cannot acquire wealth on Mars, engage in commerce with Mars, or otherwise interfere with the political, cultural, economic, and social development of the martian state.
        \item Earthlings may conduct scientific exploration on Mars only when authorized by the martian state. All scientific activities must be exclusively for peaceful purposes and must include participation or representation by martian scientists. Sharing of research and information between Mars and Earth is permitted only to pursue mutual scientific or educational goals.
        \item The use of land on Mars will be determined exclusively by the citizens of Mars. No earthlings may own, use, occupy, or otherwise lay claim to land on Mars.
        \item Any technology, resources, or other objects brought from Earth to Mars are considered gifts and become permanent fixtures of Mars. Earthlings may not make any demands for resources from martians. \citep[][pp 189-190]{haqq2022sovereign}
    \end{enumerate}
\end{quote}
The purpose of these conditions is to enable Mars to retain its sovereignty without interference from entities on Earth\revised{, in an effort to maximize the transformative outcomes of experiments in civilization on Mars}. (Note that requirement 3 refers to the ``martian state,'' which is taken in contrast to the ``earthling state'' as two planets that are juridical peers; this does not imply a single planetary government on either Mars or Earth.) In principle, the requirements for a sovereign Mars could be established within the confines of existing international law, although actually implementing such a policy would raise numerous challenges \citep[see][pp 191-195]{haqq2022sovereign} and would require a benefactor with long-term altruistic ambitions \citep{haqq2019can}. 

The framework for a sovereign Mars does not specify further details for constructing a model for governance, and numerous potential models could remain consistent with these five requirements. This paper presents one such possibility by developing a model of economic freedom that could conceptually be implemented on a sovereign Mars. The remaining discussion will refer to the five requirements for a sovereign Mars by number in order to demonstrate the possibility of a sustainable and equitable economic system that remains largely detached from the terrestrial economic system. \revised{A martian economy that remains independent could experiment with novel economic systems that may not be possible if the economies of Earth and Mars were intertwined, and the model of economic freedom presented in this paper is one such example that may be more easily implemented on a sovereign Mars.}

\revised{It is also worth noting that the idealistic framework of a sovereign Mars and the economic model developed in this paper could also be adapted for other pragmatic scenarios. It is indeed much more likely that the economies of Earth and Mars will remain combined, and any separation into separate planetary-scale economies would likely occur as a much later stage of civilizational development on Mars, rather than at the establishment of the first martian settlements. A combined Earth-Mars economy may even enable greater developmental possibilities on Mars, given that the Earth's economy will be much stronger than that on Mars at the outset, if not indefinitely. In a pragmatic sense, it could even be argued that the idea of restricting the acquisition of wealth to a specific planet or the separation of the two economies could undermine the notion of economic freedom. The sovereign Mars model is an approach to artificially separate civilizational development on Earth and Mars for the purpose of maximizing the transformative possibilities on Mars, which could even include a re-establishment of economic interactions between Earth and Mars at a much later stage of development. But the specific ideas in this paper need not be limited to the case of a sovereign Mars, and aspects or variants of this model of economic freedom could be attempted on Mars (or Earth) without necessarily requiring economic separation.}

\section{Full Reserve Banking on Mars}\label{sec:banking}

The model of economic freedom in this paper begins by addressing the problem of ensuring stability in the monetary system. \revised{The present-day monetary system allows banks to lend out much more cash than they hold in reserves, which is known as a ``fractional reserve'' system. A fractional reserve system is subject to a ``run on the bank,'' which can occur when too many depositors demand their cash at the same time, only to find that the bank cannot fulfill all its obligations. The inherent instabilities of the fractional reserve system were the motivation for} the ``Chicago Plan'' for monetary reform that was developed by several economists during the peak of the Great Depression \citep{douglas1939program}. The Chicago Plan argued that the lack of an effective monetary authority and the widespread use of fractional reserves in banking were a major driver of the Great Depression:
\begin{quote}
    A chief loose screw in our present American money and banking system is the requirement of only fractional reserves behind demand deposits. Fractional reserves give our thousands of commercial banks power to increase or decrease the volume of our circulating medium by increasing or decreasing bank loans and investments. The banks thus exercise what has always, and justly, been considered a prerogative of sovereign power. As each bank exercises this power independently without any centralized control, the resulting changes in the volume of the circulating medium are largely haphazard. This situation is a most important factor in booms and depressions. \citep[][p 20]{douglas1939program}
\end{quote}
This fractional reserve system continues today, and recent financial crises have similarly been instigated by the lack of full reserves behind bank deposits:
\begin{quote}
    If, today, those who think they have money in the bank should all ask for it, they would, of course, quickly find that the money is not there and that the banks could not meet their obligations. With 100\% reserves, however, the money \textit{would} be there; and honestly run banks could never go bankrupt as the result of a run on demand deposits. \citep[][p 22]{douglas1939program}
\end{quote}
The Chicago Plan calls for a 100\% reserve system, as this would prevent such collapses of financial institutions. The Chicago Plan also included separating the duties of a monetary authority from government influence so that the volume of currency can be adjusted in proportion with population, with the goal of maintaining constant purchasing power of the currency.

The major challenges and criticisms of the Chicago Plan are in its implementation on Earth today. Although \citet{douglas1939program} suggested a path toward implementation, and even argued that the profession of banking could remain profitable with 100\% reserves, any attempt at implementing the Chicago Plan would at the very least meet resistance from existing stakeholders that would lose their sources of wealth based on fractional reserves. Such political \revised{inertia} may be difficult to overcome on Earth, although economists in more recent recessions have continued to call for reforms toward full reserve banking \cite[e.g.,][]{laina2015proposals}. But implementation would be different in the scenario of a sovereign Mars. If Mars settlement evolves in a way that is independent from such existing systems on Earth, then the development of civilization on Mars could foster novel economic experiments in full reserve banking. Space settlement itself is a long-duration ambition, so full reserve banking or a comparable sustainable monetary policy may be an inevitability for space civilization, if not on Earth.

The establishment of a full reserve banking system on a sovereign Mars could be accomplished with three basic conditions: 
\begin{enumerate}
    \item[A.] All transactions remain on Mars. No Mars currency can be used to obtain property, wealth, or other interests on Earth. (Consistent with requirement 1).
    \item[B.] Mars currency cannot be exchanged with fractional reserve Earth currency. The monetary systems of Earth and Mars remain separate so as not to influence each other. (Consistent with requirement 2).
    \item[C.] A Mars Money Authority issues currency according to changes in carrying capacity. The fixed capacity of engineered living spaces provides a limit on the per capita supply of money.
\end{enumerate}
This approach toward monetary policy on a sovereign Mars is inspired by the Chicago Plan, and the full text by \citet{douglas1939program} provides additional details regarding the profitability of banks, the prevention of corruption, and other logistical issues of implementation that need not be repeated here. Conditions A and B are intended to prevent the economic interets of Earth from interfering with the emergence of a sustainable full reserve economy on Mars; however, it remains possible that a full reserve system that develops on Earth could potentially be more amenable to direct interaction with the martian economy. Condition C is based on the recognition by \citet{douglas1939program} that a politically independent money authority is needed to maintain monetary stability, but such an authority must be guided by a well-defined objective. In the case of Mars settlement, the carrying capacity of any infrastructure is entirely determined by physical limitations, which include a set of finite resources per person plus a ``safety factor'' to provide redundant resources in case critical life support systems fail. (For example, a settlement with a capacity of 100 people may produce enough breathable air for 200-400 people using redundant systems.) The maximum population of any particular settlement can always be calculated exactly, and a Mars Money Authority could therefore precisely adjust the supply of money according to the physical capacity of the infrastructure.

Under this model, multiple Mars currencies could co-exist, each with its own independent Mars Money Authority. Exchange between full reserve martian currencies would pose no major problems. Digital implementation of martian currencies could also solve many of the logistical problems that were unforeseen at the time the Chicago Plan was written; however, digital currency is not necessarily a requirement for this model. The adoption of such a full reserve monetary system would eliminate a major source of risk to the martian economy and promote its long-term sustainability.

\section{Diffusion of Capital Ownership on Mars}\label{sec:capital}

The second component of the model of economic freedom in this paper addresses the problem of inequality in the distribution of resources. This concept is based on the proposal developed by \citet{kelsoadler} \revised{in the book \textit{The Capitalist Manifesto}. Written} during the advent of the Cold War\revised{, the book describes a market-driven approach toward} the widespread dissemination of capital ownership, \revised{in contrast to mandatory} wealth redistribution, to \revised{enable} a good life for \revised{the greatest number}. \citet{kelsoadler} recognized that attempts at implementing the revolutionary ideas of Marx and Lenin by distributing \textit{wealth} tended to lead to authoritarian systems that undermined individual autonomy, so they instead suggested an alternative proposal for economic reform based on the distribution of wealth-producing \textit{capital}. The primary difference in these approaches is that attempts at mandating redistribution of wealth tend to enhance the centralization of government and create opportunities for authoritarianism, whereas the broad distribution of capital ownership would ensure that the production of wealth itself remains diffuse and decentralized. 

The concept of  \citet{kelsoadler} for promoting economic freedom through the diffusion of capital ownership follows from three principles of economic justice:
\begin{quote}
    The Principle of Distribution: Among those who participate in the production of wealth, each should receive a share that is proportionate to the value of the contribution each has made to the production of that wealth. \\\\
    The Principle of Participation: Every man has a natural right to life, in consequence whereof he has the right to maintain and preserve his life by all rightful means, including the right to obtain his subsistence by producing wealth or by participating in the production of it. \\\\
    The Principle of Limitation: Since everyone has a right to property in the means of production sufficient for earning a living, no one has a right to so extensive an ownership of the means of production that it excludes others from the opportunity to participate in production to an extent capable of earning for themselves a viable income; and, consequently, the ownership of productive property by an individual or household must not be allowed to increase to the point where it can injure others by excluding them from the opportunity to earn a viable income. \citep[][pp 67-68]{kelsoadler}   
\end{quote}
\revised{These principles of justice are a capitalistic (but not \textit{laissez-faire}) approach toward the distribution of wealth. The share of wealth earned by the Principle of Distribution is determined by market conditions; the wage rate of some tasks may differ from others, and the wage rates may change based on the supply and demand of the workforce or consumers. The Principle of Participation means that nobody can arbitrarily be excluded from attempting to seek employment, as work is a necessarily condition for living. And the Principle of Limitation is a prohibition against monopolies in the strongest sense---although \citet{kelsoadler} also argue for personal ethical restraint in recognizing when ``enough is enough'' (but they do not advocate for laws to make such restraints compulsory). Aspects of all these principles are included in most capitalist economies today.}

Based on these principles of justice, \citet{kelsoadler} described a justification for their proposal for diffusing capital ownership, summarized briefly here. Beginning with the Aristotelian idea that sufficient wealth is needed for living a good life, the means of generating wealth is some combination of labor (i.e., wealth generated by performing physical work) and capital (i.e., wealth generated by owning property, such as dividends, rents, or royalties). If capital ownership is then diffused as widely as possible, rather than concentrated in the hands of a few, then most people will be able to generate enough wealth through a combination of labor and capital to live a good life. Capital owners would continue to receive benefits of property ownership, and workers would be fairly compensated according to market conditions. The diffusion of capital ownership would also be facilitated by technological advances, which would reduce the required amount of labor per person and also lower the costs of living a good life. 

The widespread diffusion of capital ownership would improve the quality of life for \revised{nearly} everyone, but \citet{kelsoadler} noted that such an outcome could not occur in a mixed government that tends toward heavy regulations, which paradoxically inhibits the diffusion of capital ownership rather than encourages it. But the authors also specify that their proposal is ``far from being a system of \textit{laissez-faire}'' \citep[][pp 153]{kelsoadler} and that government regulation would be needed to maintain the three principles of economic justice. \revised{For example, capital ownership would include the responsibility of managing the capital (whether this is looking after physical infrastructure, managing financial assets, or other tasks), and some capital owners may be unable to successfully manage their capital due to ill fortune or lack of abilities. In such cases, some individuals may lose their capital and depend on wage earnings entirely, even if a majority of citizens remain capital owners. Some individuals who cannot work or manage capital may even require limited forms of financial assistance, ideally provided through charities but as a function of government if necessary. Another example is the role of government to uphold the Principle of Limitation through anti-trust laws that would unfairly exclude others from competition.} \citet{kelsoadler} discussed many of the objections to their proposal as well as specific ideas for how to stimulate the diffusion of capital ownership on Earth at the time of their writing, but such details need not be repeated here. 

Implementing the diffuse ownership of capital on Earth would remain challenging given the likely resistance by holders of large capital stakes, but this approach could conceptually be applied on a sovereign Mars with two basic conditions:
\begin{enumerate}
    \item[D.] All citizens of Mars are capital owners. All martians obtain wealth from a combination of capital ownership and labor. (Consistent with requirement 4). 
    \item[E.] All fixtures of Mars become co-owned capital. The diffusion of new capital will be determined by the citizens of Mars. (Consistent with requirement 5).
\end{enumerate}
This approach toward the diffusion of capital ownership on Mars would prevent the emergence of central authoritarian entities from undermining the individual liberty or economic freedom of people living on a sovereign Mars. The widespread ownership of martian infrastructure would also reduce the risk of any critical life support technology from being monopolized by a small number of actors: for example, the widely diffused ownership of air production facilities would only serve to promote affordable pricing for air, rather than encourage poor air management practices \citep[c.f.,][]{stevens2015price}. 

New capital ownership opportunities would also arise with the expansion of any space settlement infrastructure, with the maximum number of capital owners limited by the total carrying capacity and safety factors of any infrastructure. This means that any new citizens of Mars (through birth or immigration) would acquire capital upon being granted citizenship, along with being granted the use or ownership of sufficient infrastructure for living. The specific details for how such capital distribution should be accomplished would be determined by the citizens of Mars, but the general concept of widely diffusing capital to enable a large number of citizens to benefit from wealth generation remains a viable foundation for organizing an economy on a sovereign Mars.

\section{Tourism on Mars}\label{sec:tourism}

The conditions for a sovereign Mars are intended to minimize the influence of Earth on Mars by limiting the interaction between the two planets, at least during the time when civilization on Mars is developing. But this model for economic freedom can also include a limited form of tourism on Mars, which could provide a way to accelerate the development of capital infrastructure on Mars---and meet the demands of earthling tourists---but without undermining the sovereignty or autonomy of martian citizens. 

Implementing tourism on a sovereign Mars could be accomplished with three basic conditions:
\begin{enumerate}
    \item[F.] Tourists cannot own capital on Mars. (Consistent with requirement 2). 
    \item[G.] Tourists cannot bring Mars currency back to Earth. (Consistent with requirement 1 and condition A).
    \item[H.] Tourists can obtain services on Mars for the exchange of real goods only. Tourists cannot conduct currency transactions on Mars. (Consistent with requirement 2 and condition B).
\end{enumerate}
This approach is intended to allow tourism to exist in a way that can be limited by the citizens of Mars as needed. Conditions F and G are intended to prevent tourists from acquiring economic interests on Mars or from compromising the full reserve monetary system on Mars. Condition H enables the possibility for tourists to engage in a form of bartering, using luxury goods or other supplies brought from Earth. The exchange value for such goods would be valued in Mars currency, but no exchange of currency would take place. All tourism capital would be fixtures of Mars (even if constructed by entities from Earth), to be co-owned and managed by martians. 

Under this model, a commercial tourist experience could still be facilitated by an Earth-based company. One scenario could involve an Earth-based tourism company that sells an all-inclusive Mars experience, which includes round-trip travel plus all other expenses for a price paid in Earth currency. The tourism company would pay for the expenses of the launch using Earth currency and would carry enough Earth goods to exchange for lodging, food, souvenirs, and other services on Mars for all their clients. This would allow Earth tourists to experience a short-duration visit to Mars but without interfering with the stability of the Mars monetary system or distribution of capital ownership. The introduction of new goods from Earth as barter for services on Mars would not change the supply of money on Mars but would cause prices on Mars to decrease. The extent of such tourism would be determined by the citizens of Mars, and safeguards should be implemented to prevent indirect influence of development of a sovereign Mars by Earth-based entities. Nevertheless, such an exception to interactions between Earth and Mars could enable the exchange of goods and services between two planetary economies.

\section{Conclusion}

This model for economic freedom has focused on implementation within the framework of a sovereign Mars. The conditions of a sovereign Mars provide not only a potential framework for conceptualizing systems of governance, but the value of these conditions as a thought experiment can promote novel ideas that may not otherwise have emerged. The resulting model based on a full reserve monetary system and diffusion of capital ownership does not necessarily need to be constrained to the conditions of a sovereign Mars, and this model could be adapted in whole or in part for other feasible approaches to space settlement, if not on Earth. 

The economic model in this paper is admittedly idealistic, and even if its implementation is conceptually feasible, it may be unlikely to actually be adopted in the future on Mars or Earth. But idealistic models can still suggest features for implementation in reality, and the application of idealistic models to the martian environment can provide relevant insights into the use of more realistic tools on Earth. For example, monetary policy on Mars is somewhat easier because the carrying capacity is exactly constrained by the physical infrastructure; Earth's carrying capacity may be more difficult to quantify, but this capacity is limited as well. Creative problem solving on Earth can benefit from learning to ``think like a martian'' in order to identify physically-driven limits that may be less apparent from the perspective of Earth. 

An important lesson from the economic model in this paper is that diffuse capital ownership and limited government can enable freedom in space. Centralized control through a world space agency or other new sovereign entities is not necessarily the only way to curtail the concentration of power in space, and numerous options should be explored as possibilities for effective governance of the space economy.

\section*{Acknowledgments}
This research did not receive any specific grant from funding agencies in the public, commercial, or not-for-profit sectors. Any opinions, findings, and conclusions or recommendations expressed in this material are those of the author and do not necessarily reflect the views of any employer.

\bibliographystyle{elsarticle-num-names} 
\bibliography{refs}

\begin{thebibliography}{15}
\expandafter\ifx\csname natexlab\endcsname\relax\def\natexlab#1{#1}\fi
\providecommand{\url}[1]{\texttt{#1}}
\providecommand{\href}[2]{#2}
\providecommand{\path}[1]{#1}
\providecommand{\DOIprefix}{doi:}
\providecommand{\ArXivprefix}{arXiv:}
\providecommand{\URLprefix}{URL: }
\providecommand{\Pubmedprefix}{pmid:}
\providecommand{\doi}[1]{\href{http://dx.doi.org/#1}{\path{#1}}}
\providecommand{\Pubmed}[1]{\href{pmid:#1}{\path{#1}}}
\providecommand{\bibinfo}[2]{#2}
\ifx\xfnm\relax \def\xfnm[#1]{\unskip,\space#1}\fi
\bibitem[{Milligan(2014)}]{milligan2014nobody}
\bibinfo{author}{T.~Milligan}, \bibinfo{title}{Nobody owns the moon: The ethics of space exploitation}, \bibinfo{publisher}{McFarland}, \bibinfo{year}{2014}.
\bibitem[{Szocik(2019)}]{szocik2019human}
\bibinfo{author}{K.~Szocik},
\newblock \bibinfo{title}{The human factor in a mission to mars},
\newblock \bibinfo{journal}{An Interdisciplinary Approach}  (\bibinfo{year}{2019}).
\bibitem[{Gilley(2020)}]{gilley2020space}
\bibinfo{author}{J.~Gilley}, \bibinfo{title}{Space civilization: An inquiry into the social questions for humans living in space}, \bibinfo{publisher}{Lexington Books}, \bibinfo{year}{2020}.
\bibitem[{Froehlich(2021)}]{froehlich2021assessing}
\bibinfo{author}{A.~Froehlich}, \bibinfo{title}{Assessing a Mars agreement including human settlements}, \bibinfo{publisher}{Springer}, \bibinfo{year}{2021}.
\bibitem[{Cockell(2022{\natexlab{a}})}]{cockell2022institutions}
\bibinfo{author}{C.~S. Cockell}, \bibinfo{title}{The Institutions of Extraterrestrial Liberty}, \bibinfo{publisher}{Oxford University Press}, \bibinfo{year}{2022}{\natexlab{a}}.
\bibitem[{Cockell(2022{\natexlab{b}})}]{cockell2022interplanetary}
\bibinfo{author}{C.~S. Cockell}, \bibinfo{title}{Interplanetary Liberty: Building Free Societies in the Cosmos}, \bibinfo{publisher}{Oxford University Press}, \bibinfo{year}{2022}{\natexlab{b}}.
\bibitem[{Haqq-Misra(2022)}]{haqq2022sovereign}
\bibinfo{author}{J.~Haqq-Misra}, \bibinfo{title}{Sovereign Mars: Transforming Our Values through Space Settlement}, \bibinfo{publisher}{University Press of Kansas}, \bibinfo{year}{2022}.
\bibitem[{Nesvold(2023)}]{nesvold2023off}
\bibinfo{author}{E.~Nesvold}, \bibinfo{title}{Off-Earth: Ethical questions and quandaries for living in Outer Space}, \bibinfo{publisher}{MIT Press}, \bibinfo{year}{2023}.
\bibitem[{Schwartz et~al.(2023)Schwartz, Billings, and Nesvold}]{schwartz2023reclaiming}
\bibinfo{author}{J.~S. Schwartz}, \bibinfo{author}{L.~Billings}, \bibinfo{author}{E.~Nesvold}, \bibinfo{title}{Reclaiming Space: Progressive and Multicultural Visions of Space Exploration}, \bibinfo{publisher}{Oxford University Press}, \bibinfo{year}{2023}.
\bibitem[{Haqq-Misra(2023)}]{haqq2023constraints}
\bibinfo{author}{J.~Haqq-Misra},
\newblock \bibinfo{title}{Constraints on interstellar sovereignty},
\newblock \bibinfo{journal}{arXiv preprint arXiv:2308.11076}  (\bibinfo{year}{2023}).
\bibitem[{Douglas et~al.(1939)Douglas, Fisher, Graham, Hamilton, King, and Whittlesey}]{douglas1939program}
\bibinfo{author}{P.~H. Douglas}, \bibinfo{author}{I.~Fisher}, \bibinfo{author}{F.~D. Graham}, \bibinfo{author}{E.~J. Hamilton}, \bibinfo{author}{W.~I. King}, \bibinfo{author}{C.~R. Whittlesey}, \bibinfo{title}{A program for monetary reform}, \bibinfo{year}{1939}.
\bibitem[{Kelso and Adler(1958)}]{kelsoadler}
\bibinfo{author}{L.~O. Kelso}, \bibinfo{author}{M.~J. Adler}, \bibinfo{title}{The Capitalist Manifesto}, \bibinfo{publisher}{Random House}, \bibinfo{year}{1958}.
\bibitem[{Haqq-Misra(2019)}]{haqq2019can}
\bibinfo{author}{J.~Haqq-Misra},
\newblock \bibinfo{title}{Can deep altruism sustain space settlement?},
\newblock \bibinfo{journal}{The human factor in a mission to mars: An interdisciplinary approach}  (\bibinfo{year}{2019}) \bibinfo{pages}{145--155}.
\bibitem[{Lain{\`a}(2015)}]{laina2015proposals}
\bibinfo{author}{P.~Lain{\`a}},
\newblock \bibinfo{title}{Proposals for full-reserve banking: A historical survey from david ricardo to martin wolf},
\newblock \bibinfo{journal}{Economic Thought} \bibinfo{volume}{4} (\bibinfo{year}{2015}) \bibinfo{pages}{1--19}.
\bibitem[{Stevens(2015)}]{stevens2015price}
\bibinfo{author}{A.~H. Stevens},
\newblock \bibinfo{title}{The price of air},
\newblock in: \bibinfo{editor}{C.~S. Cockell} (Ed.), \bibinfo{booktitle}{Human Governance Beyond Earth: Implications for Freedom}, \bibinfo{publisher}{Springer}, \bibinfo{year}{2015}, pp. \bibinfo{pages}{51--61}.

\end{thebibliography}

\end{document}